\documentclass[twocolumn,showpacs,amsmath,amssymb,superscriptaddress,aps,prl]{revtex4-1}

\usepackage{graphicx}
\usepackage{dcolumn}
\usepackage{bm}
\usepackage{xcolor}

\newcommand{\be}{\begin{equation}}
\newcommand{\ee}{\end{equation}}
\newcommand{\bea}{\begin{eqnarray}}
\newcommand{\eea}{\end{eqnarray}}
\newcommand{\la}{\left\langle}
\newcommand{\ra}{\right\rangle}

\begin{document}

\title{Nonequilibrium information erasure below kTln2}

\author{Michael Konopik}
\affiliation{Department of Physics, Friedrich-Alexander-Universit\"at Erlangen-N\"urnberg, D-91058 Erlangen, Germany}
\affiliation{Institute for Theoretical Physics I, University of Stuttgart, D-70550 Stuttgart, Germany}
\author{Alexander Friedenberger}
\affiliation{Department of Physics, Friedrich-Alexander-Universit\"at Erlangen-N\"urnberg, D-91058 Erlangen, Germany}
\author{Nikolai Kiesel}
\affiliation{Vienna Center for Quantum Science and Technology (VCQ), Faculty of Physics, University of Vienna, A-1090 Vienna, Austria}
\author{Eric Lutz}
\affiliation{Department of Physics, Friedrich-Alexander-Universit\"at Erlangen-N\"urnberg, D-91058 Erlangen, Germany}
\affiliation{Institute for Theoretical Physics I, University of Stuttgart, D-70550 Stuttgart, Germany}

\begin{abstract}
Landauer's principle states that information erasure requires heat dissipation. Landauer's  original result focused on  equilibrium memories. We here investigate the reset of information stored in a nonequilibrium state of a symmetric two-state memory. We derive a nonequilibrium generalization of the erasure principle and demonstrate that the corresponding bounds on work and heat may be  reduced to zero. We further introduce reset  protocols that harness the initial preparation energy and entropy and so allow to reach  these nonequilibrium  bounds. We finally provide  numerical simulations with realistic parameters of an optically levitated nanosphere memory that support these findings. Our  results indicate that local dissipation-free information reset is possible away from equilibrium. \end{abstract}


\maketitle

According to the standard (equilibrium) formulation of Landauer's principle, the erasure  of one bit of information generates  at least $kT \ln 2$ of heat and consumes the same amount of work \cite{lan61}. Here $T$ is the temperature of the environment to which the memory  device is coupled, $k$ the Boltzmann constant and $\ln2$ the information content. Information erasure is thus    unavoidably  dissipative \cite{ple01,mar09}. Landauer's principle is a central result of the thermodynamics of information  that applies to all logically irreversible transformations \cite{lut15,par15}. It additionally imposes a fundamental physical limit to the downsizing of binary switches, such as field effect transistors  \cite{fra02,pop10,the17}.   The existence of the Landauer bound has been established in   a number of  experiments in which two-state memories have been realized with an optical tweezer \cite{ber12}, an electrical circuit \cite{orl12}, a feedback trap \cite{jun14} and nanomagnets \cite{mar16,hon16}. Meanwhile, growing energy consumption and dissipation in modern integrated electronics  has become  a major technological challenge that threatens future progress \cite{fra02,pop10,the17}.  It has recently been shown  that the work required for erasure may be reduced to zero in nonequilibrium asymmetric memories in the overdamped regime \cite{sag09,dil10,gav17}. We here focus on the more pressing issue of control and suppression of heat dissipation.

The study of nonequilibrium memories is not purely academic. Two different types of electronic storage devices are   usually distinguished \cite{sta15}. Read-only-memories (ROM) (and their variants EPROM and EEPROM \cite{com1}) are  non-volatile memories that retain the information stored in them in the absence of a power source. Information is here  encoded  in an equilibrium state, as considered in Landauer's original  principle \cite{lan61}. By contrast,  random-access memories (RAM), the most common memories in modern computers, are  volatile and  the stored data is lost when  power is switched off. Information is in this case encoded in a nonequilibrium  state, whose preparation  requires a given  amount of energy and entropy. At the same time,   novel switching devices, beyond the standard FET technology, are currently being explored in order to decrease power dissipation \cite{the10}. Promising examples include tunable nanomechanical oscillators that operate in the weakly damped regime \cite{mah08,bag11,ven13,ric17}.

Motivated by these observations, we here perform a detailed investigation of  nonequilibrium information erasure. We first derive a  generalization of Landauer's principle that holds for information initially stored in a nonequilibrium state. We show that  both work \textit{and} heat associated with the reset process  may be theoretically  reduced below the equilibrium Landauer bounds of $kT \ln 2$, provided the preparation energy and entropy are properly harnessed. Both quantities may even change sign, indicating that work may be produced and heat absorbed  with the help of the prepared initial state.
We stress that these findings do not violate the second law of thermodynamics, but directly follow from it when applied to the considered nonequilibrium situation. We further introduce novel erasure protocols that allow to reach these nonequilibrium bounds in a generic, symmetric double-well potential. We finally discuss a possible experimental verification of the nonequilibrium erasure principle in the underdamped regime using an optically levitated nanosphere \cite{asp14} and provide extensive numerical simulations of the  process with realistic parameters.

\textit{Nonequilibrium erasure principle.}  We begin by analyzing an erasure cycle that consists of a preparation and a reset phase. To that end, we consider a general memory device weakly coupled to a heat bath at temperature $T$. The total entropy change for system and bath during the reset phase is  $\Delta S_\text{res} = \Delta S_\text{mem} + \Delta S_\text{bat} \geq 0$ \cite{zem97}. Owing to its large size, the bath always remains in equilibrium and thus $\Delta S_\text{bat} = Q/T$, where $Q$ is the heat dissipated into the environment. The system is assumed to be initially in a nonequilibrium state  with  phase-space distribution $\rho(x,p,0)$ where $x$ denotes the position and $p$ the momentum. After reset of duration $\tau$, the memory is in   state $\rho(x,p,\tau)$. The work done on the system during reset is  $W= \Delta F_\text{eq} + T \Delta I + T \Delta S_\text{res}$ \cite{tak10,def11,esp11},
 where $\Delta F_\text{eq}$ is the equilibrium free energy difference and $I(t)= S(\rho(t)||\rho_\text{eq}(t))= \int dxdp \,\rho(t) \ln[\rho(t)/\rho_\text{eq}(t)]$  the relative entropy between the nonequilibrium state $\rho(t)$ and the corresponding equilibrium state $\rho_\text{eq}(t)$. The entropic distance $I$  may be interpreted as the amount of information needed to prepare $\rho(t)$ from  $\rho_\text{eq}(t)$ \cite{sch80}.

Using the first law, the dissipated heat  may   next be written as $Q = W - \Delta U_\text{res} = \Delta F_\text{eq} +T \Delta I + T \Delta S_\text{res} - \Delta U_\text{res}$, where  $\Delta U_\text{res}$ is the variation of internal energy during reset.  For a complete erasure cycle consisting of preparation and reset, we have $\Delta U = \Delta U_\text{pre} +  \Delta U_\text{res}$ = 0. For an equilibrium memory,  $\Delta U_\text{pre} =  \Delta U_\text{res} =\Delta I = 0$. In addition, for the reset of one bit of information, $\Delta F_{\text{eq}} = - T \Delta S_{\text{mem}} = k T\ln 2$, with the Gibbs-Shannon entropy of the memory $S_\text{mem}(t) = -k \int dx dp\, \rho(t) \ln \rho(t)$. According to the second law, the total entropy production is positive, $\Delta S_\text{res}\geq 0$ \cite{zem97}.  As a result, we obtain the following two inequalities for heat and work,
 \bea
Q &\geq& Q_L= kT \ln 2 + T [I(\tau) - I(0)] - \Delta U_\text{res}, \label{1} \\
W &\geq& W_L= kT \ln 2 + T [I(\tau) - I(0)]. \label{2} 
 \eea
 \begin{figure}[t]
\includegraphics[width=0.45\textwidth]{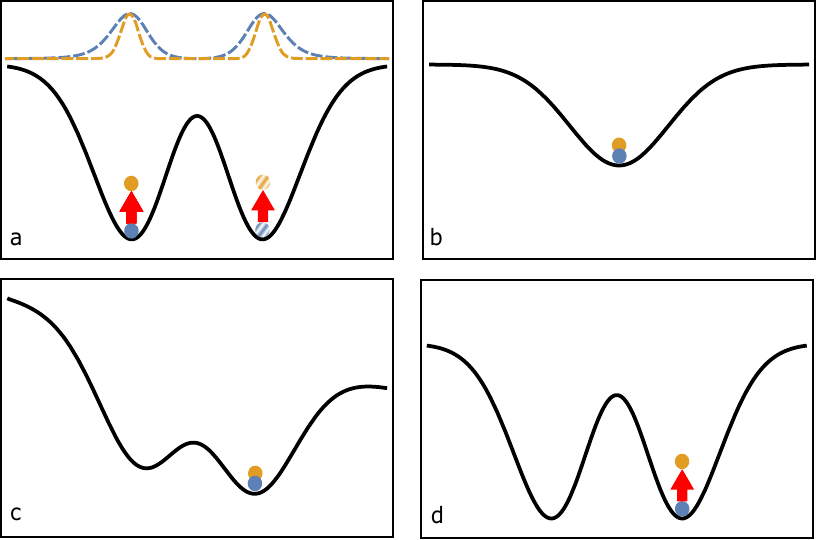}
\caption{Two-state memory. Double-well potential \eqref{4} used as a generic  symmetric two-state memory. (a) Initially, 1 bit of information is stored  in a configuration where the wells are occupied with probability 1/2. Blue (yellow) dashed lines represent  equilibrium (nonequilibrium)  distributions used for storage. (b-d) Information is erased  by bringing the particle with probability 1 to  the right well by cyclically lowering the barrier   and applying a tilt.  Blue (yellow) particles depict equilibrium (nonequilibrium) states. The red arrows schematically show the preparation of the nonequilibrium state. } 
\label{f1}
\end{figure}
 Equations \eqref{1}-\eqref{2} are nonequilibrium generalizations of Landauer's  erasure principle to which they reduce for initial and final equilibrium states that correspond to  $I(0) = I(\tau) = 0$ and $U(0) = U(\tau)$. We note that Eqs.~\eqref{1}-\eqref{2} may be compactly written as  $Q \geq \Delta {\cal F} -\Delta U_\text{res}$ and $W \geq \Delta {\cal F}$, where ${\cal F} = F + T I$ is the  information free energy \cite{par15,def12}. 
We observe  that the nonequilibrium Landauer bounds for heat and work, $Q_L$ and $W_L$,  can in principle be controlled  through the initial entropic distance to equilibrium $I(0)$ and the preparation energy, $\Delta U_\text{pre} =  -\Delta U_\text{res}$.  This opens the fascinating possibility to reduce both the amounts of heat and work required for reset below the equilibrium value of $kT \ln 2$. The essential questions that we here address are (a) whether the nonequilibrium Landauer bounds \eqref{1}-\eqref{2}  can be actually reached in practice and (b) if yes, how?

 To answer these questions, we investigate a Brownian particle  in a symmetric double-well potential. Such a two-state system may be regarded as a generic model for an elementary memory  and has been employed in the experiments \cite{ber12,jun14,mar16,hon16}. We describe the dynamics of the particle with  the underdamped Langevin equation \cite{ris89},  
\begin{equation}
\label{3}
m\ddot x +\gamma \dot x + V'(x,t) -  A f(t) = F(t), 
\end{equation}
where $m$ is the mass of the particle,  $\gamma$  the friction coefficient, $f(t)$ a tilting force with amplitude $A$, and $F(t)$ a centered white noise force with variance $\la F(t)F(t')\ra= 2m\gamma kT \delta(t-t')$. For concreteness, the symmetric double-well potential $V(x,t)$  is taken to  be of the form (Fig.~1),
\begin{equation}
\label{4}
V(x,t) = -\left[h(t) a+g(t)\frac{b}{2} x^2\right] \exp\left(\frac{-c x^2}{2}\right),
\end{equation}
with a tunable barrier height via $g(t)$ and a tunable barrier width via the  function $h(t)$.
Such  a potential appears naturally in the  optomechanical setup  discussed below. We stress that our findings do not depend on the specific shape of the double-well potential considered. 

\begin{figure}[t]
\includegraphics[width=0.44\textwidth]{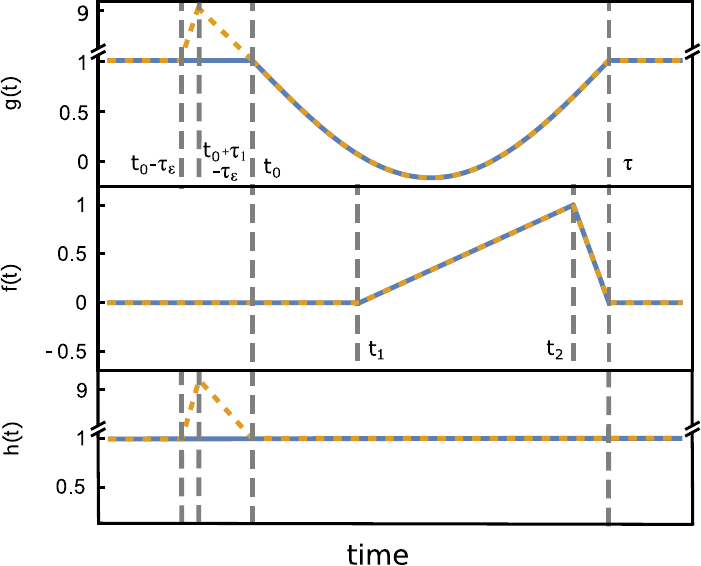}
\caption{Reset protocols. The sawtooth function $f(t)$, Eq.~\eqref{5}, applies the tilt toward the right well, while the function $g(t)$, Eq.~\eqref{6}, cyclically modulates the barrier height. The barrier width is controlled by $h(t)$. The blue (continuous) lines show the equilibrium protocol (5)-(6) and the yellow (dashed) lines the nonequilibrium protocol (9)-(10) designed to harness the nonequilibrium preparation energy and entropy.} 
\label{f2}
\end{figure}

\textit{Equilibrium erasure protocol.} To put our nonequilibrium results into proper perspective, we first investigate commonly used equilibrium erasure protocols \cite{ber12,jun14,mar16,hon16}. The particle is initially prepared to occupy either of the two wells with equal probability $1/2$ (Fig.~1).  In this configuration $S_\text{mem}(0)= k\ln 2$ and the memory stores one bit of information. That information is erased by modulating the shape of the confining potential during time $\tau$ such that the particle ends up with probability 1 in one of the wells, $S_\text{mem}(\tau)= 0$, irrespective of its initial location \cite{ben82}. The reset operation is implemented by decreasing the height of the barrier via $g(t)$ and applying the tilt $f(t)$ in a cyclic manner, $h(t) = 1$ throughout this process (Fig.~2, blue continuous lines) \cite{dil09}:
\bea
\label{5}
f(t) &=&\begin{cases} 
	{(t-t_1)}/{(t_2-t_1)},&\text{} t_1<t\leq t_2\\
	1-({t-t_2})/({\tau-t_2}),&\text{} t_2<t \leq \tau\\
		0,& \text{otherwise}  \\
	\end{cases}\\
	g(t) &=&\begin{cases} 
	1- B \sin\left[\frac{\pi (t-t_0)}{\tau-t_0}\right],&\text{} t_0<t \leq \tau\\
		1,& \,\text{otherwise}  \\
	\end{cases}\label{6}
	\eea
The parameter $B$ controls the amplitude of the barrier lowering. Erasure protocols of this type have been implemented in the recent experiments \cite{ber12,jun14,mar16,hon16}, where information was encoded in an initial equilibrium state. We therefore call them equilibrium erasure protocols.

\begin{figure}[t]
\includegraphics[width=0.38\textwidth]{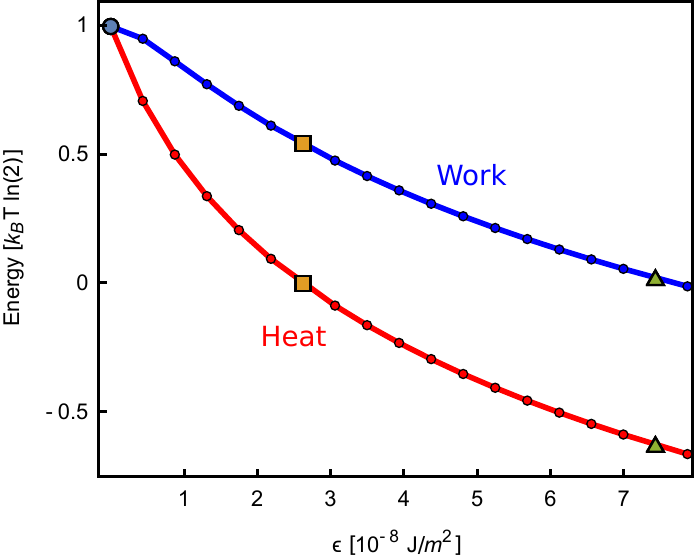}
\caption{Nonequilibrium Landauer bounds. Heat $Q_L$ (red), Eq.~\eqref{1}, and work $W_L$ (blue), Eq.~\eqref{2}, as a function of the parameter $\varepsilon$ that quantifies the departure from equilibrium of the initial  state, Eq.~\eqref{7}. Both nonequilibrium bounds decrease and  become negative with increasing values of $\varepsilon$.} 
\label{fig3}
\end{figure} 

We choose  the initial nonequilibrium distribution of the symmetric memory device to be given by (Fig~1),
\be
\label{7}
\rho(x,p,0) = \frac{1}{Z'_{b+\varepsilon}} \exp\left[-\beta \left(\frac{p^2}{2m}+ V'_{b+\varepsilon}(x)\right)\right].
\ee
where $Z'_{b+\varepsilon}$  is the normalization constant and $\beta=1/(kT)$  the inverse temperature. The  modified potential reads,
\begin{equation}
\label{7a}
V'_{b+\varepsilon}(x) = -\left[ a'(b+\varepsilon)+\frac{b+\varepsilon}{2} x^2\right] \exp\left(-\frac{c x^2}{2}\right),
\end{equation}
with $a'(b') = b'/(2c)(2- c \bar x^2)$ and $\bar x = \sqrt{2/(bc)(b-ca)}$ the fixed position of the potential minima (Fig. 1). Equations \eqref{7}-\eqref{7a} are chosen in order  to decrease  energy and entropy  of the initial nonequilibrium state as  compared to the equilibrium state. In particular, the nonequilibrium state is narrower than the corresponding equilibrium state (Fig.~1). The parameter $\varepsilon$ controls the departure from equilibrium and Eq.~\eqref{7} reduces to the equilibrium distribution $\rho_\text{eq}(x,p,0)$ for $\varepsilon=0$.

In order to study the approach to the nonequilibrium Landauer bounds \eqref{1}-\eqref{2}, we simulate the reset process by numerically solving the Langevin equation \eqref{3} for the protocols \eqref{5} and \eqref{6} with experimentally realistic parameters,  using a 4th-order Runge-Kutta method (Supplemental Material). The starting points of the simulations are randomly generated according to $\rho(x,p,0)$, Eq.~\eqref{6}. The final distribution $\rho(x,p,\tau)$ is determined from the end points of the simulated trajectories. Figure 3 shows the Landauer bounds $Q_L$ and $W_L$, Eqs.~\eqref{1}-\eqref{2}, as a function of the parameter $\varepsilon$, for an infinitely large  $\tau$, which corresponds to a final equilibrium state with $I(\tau)=0$. As expected, both $Q_L$ and $W_L$ decrease with increasing $\varepsilon$, that is, with increasing initial departure from equilibrium $I(0)$, and eventually turn negative. The actual work and heat, $W$ and $Q$, for the equilibrium reset  process are plotted  in  Figs.~4a) and 5a) as a function of the duration $\tau$ for three different values of $\varepsilon$, indicated  by the circle, square and triangle symbols in Fig.~3.   We observe that,  for all values of the parameter $\varepsilon$, work $W$  remains above the   equilibrium Landauer bound of  $kT \ln 2$ (horizontal dashed lines) in the limit of long  times, while heat $Q$ is only slightly reduced below that value. The nonequilibrium bounds \eqref{1}-\eqref{2} are thus not reached.  By naively implementing the commonly used equilibrium  protocols \eqref{5} and \eqref{6} \cite{ber12,jun14,mar16,hon16}, the extra energy and entropy that are required to prepare the initial nonequilibrium state $\rho(0)$ are simply dissipated at the beginning of the erasure process, and therefore lost.

\textit{Nonequilibrium erasure protocol.} In order to successfully harness the preparation  energy and entropy to  reduce  work and heat dissipated during reset, we modify  the  equilibrium protocol $g(t)$, Eq.~\eqref{6}, and $h(t)$ by adding an additional modulation of the potential at the start of the process (Fig.~2, yellow dashed lines):
\bea
\label{q2}
	\bar g(t) &=&\!\begin{cases} 
	1+ \frac{b'-1}{b}\frac{t'}{\tau_1}, & \!\text{}  t_0-\tau_\varepsilon < t\leq  t_0-\tau_\varepsilon + \tau_1\\
1+ \frac{b'-1}{b}\left(\frac{\tau_\varepsilon-t'}{\tau_\varepsilon-\tau_1} \right),& \!\text{} t_0- \tau_\varepsilon+\tau_1 < t\leq t_0\\	
	1- B \sin\left[\frac{\pi (t-t_0)}{\tau-t_0}\right],&\!\text{} t_0<t \leq \tau\\
		1,& \!\text{otherwise}  \\
	\end{cases}\\
\bar h(t) &=&\begin{cases} 
1+ \frac{a'-1}{a}\frac{t'}{\tau_1},& \text{}   t_0-\tau_\varepsilon < t\leq t_0-\tau_\varepsilon + \tau_1\\
1+ \frac{a'-1}{a}\left(\frac{\tau_\varepsilon-t'}{\tau_\varepsilon-\tau_1}  \right),& \text{}   t_0- \tau_\varepsilon+\tau_1 < t\leq t_0\\	
		1,& \text{otherwise}  \\
	\end{cases}	
	\label{q3}
	\eea 	
with $b'= b + \epsilon$ and $t'=t-t_0+\tau_\varepsilon$.
The constants $\tau_\varepsilon \,(\sim20 \text{ms})$ and $\tau_1 \,(\sim0.2 \text{ms})$ depend in general on the parameter $\varepsilon$. They are chosen in order to effectively extract the preparation energy and entropy of the initial nonequilibrium state.
Figures 4b) and 5b) show the erasure work $W$ and the dissipated heat $Q$ computed with the nonequilibrium erasure protocol \eqref{q2}-\eqref{q3} for the same three values of $\varepsilon$ as before (the tilting function $f(t)$, Eq.~\eqref{5}, is kept unmodified). We see that both quantities  now asymptotically approach the nonequilibrium Landauer bounds \eqref{1}-\eqref{2} for long  times. In particular, heat  vanishes for $\varepsilon = 26.24 \cdot 10^{-9} J/m^2$  and work vanishes for $\varepsilon = 74.33 \cdot 10^{-9} J/m^2$. In these cases, the memory has been reset without work and heat dissipation.

\begin{figure}[t]
\includegraphics[width=0.43\textwidth]{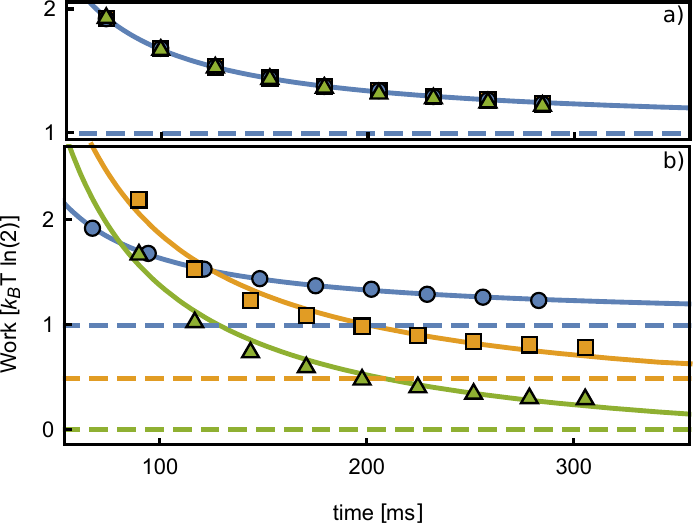}
\caption{Work consumed  during reset.  Work $W$ as a function of the reset time $\tau$ for three values of the nonequilibrium parameter $\varepsilon$: $\varepsilon=0$ (blue circles),  $\varepsilon=26.24\cdot 10^{-9}J/m^2$ (yellow squares), $\varepsilon=74.33 \cdot 10^{-9} J/m^2$  (green triangles).  (a) is obtained with the equilibrium protocols (5)-(6); (b) is obtained with the nonequilibrium protocols (9)-(10). The continuous lines display a fit with the function $\sim1/\tau$. The dashed lines show the nonequilibrium Landauer bound $W_L$ (2).} 
\label{fig4}
\end{figure}

\textit{Experimental setup.} To experimentally verify the nonequilibrium Landauer principle \eqref{1}-\eqref{2}, we propose to use an optically levitated nanoparticle. The two main advantages of this setup is that the optical confining potential can be flexibly tuned and that the underdamped regime, described by  the Langevin equation \eqref{3}, is easily accessible. 
While first experiments on optical levitation have already been realized in the early 1970ies \cite{ash76}, more recently an excellent experimental control in ultra-high vacuum has been demonstrated in optical tweezers \cite{yin13}. In order to implement a double-well potential with a controlled barrier height $g(t)$, we suggest to form an optical trap by the combination of a TEM$_{00}$ and a TEM$_{01}$ mode inside a Fabry-Perot cavity, leading to a potential of the form  \eqref{4} (alternative options to create complex potential landscapes may be found in Ref.~\cite{dho11}). This configuration ensures particularly low intensity fluctuations of the optical trap and the additional cavity power enhancement allows to use a wide trap (here approx. $80$ micrometers) with low driving powers of only a few milliwatt. Stable optical trapping directly inside an optical cavity has  been lately  successfully demonstrated in Refs.~\cite{kie13,mil15,mes15}.

The height function $g(t)$, Eqs.~\eqref{6}, (9), may be changed by varying the power of the cavity modes, while the tilt $f(t)$, Eq.~\eqref{5}, may be implemented using the radiation pressure from a cavity-independent light source. The feasibility of this approach is underlined by a similar strategy that has been recently followed to achieve radial feedback cooling of levitated microparticles  \cite{li11}. The experimental parameters chosen for the numerical simulations (summarized in Tab.~I in the Supplemental) are taken from Ref.~\cite{kie13}. Additional information on experimental details and position readout may be found in the Supplemental Material \cite{sup}.

\begin{figure}[t]
\includegraphics[width=0.43\textwidth]{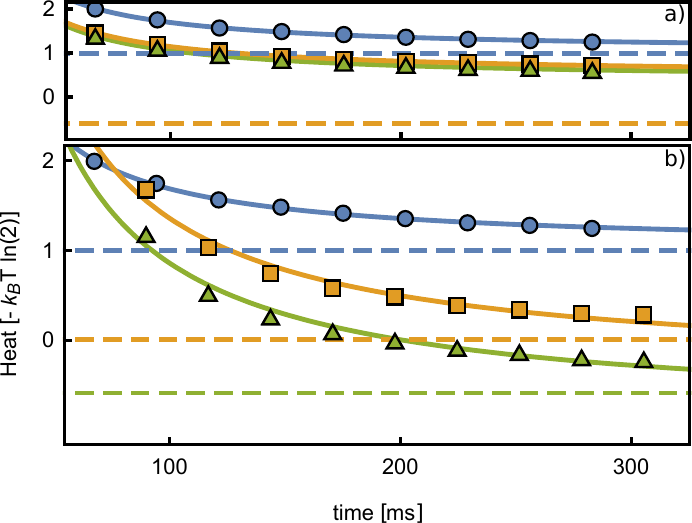}
\caption{Heat dissipated  during reset.  Heat $Q$ as a function of the reset time $\tau$ for the same three values of the nonequilibrium parameter as in Fig.~4.  (a) is obtained with the equilibrium protocols (5)-(6); (b) is obtained with the nonequilibrium protocols (9)-(10). The continuous lines display a fit with the function $\sim1/\tau$. The dashed lines show the nonequilibrium Landauer bound $Q_L$ (1).} 
\label{fig5}
\end{figure}

\textit{Discussions.} We have studied the erasure of information encoded in a nonequilibrium  state of a symmetric memory. We have concretely derived a nonequilibrium extension  of Landauer's principle  and shown that the corresponding bounds for heat and work, Eqs.~\eqref{1}-\eqref{2}, may both be reduced to zero. Using a generic model based on an underdamped Brownian particle in a double-well potential, we have demonstrated that these nonequilibrium  limits  may be reached by properly tuning the erasure protocol to harness the initial preparation energy and entropy.  We have further performed detailed numerical simulations of the nonequilibrium Landauer principle with realistic parameters using an optically levitated nanosphere to support these findings. In contrast to the standard equilibrium situation, the complete nonequilibrium erasure cycle here consists of distinct preparation and reset stages (the preparation stage being absent for equilibrium erasure).  This offers new and powerful means to control the thermodynamics of logically irreversible operations.  By, for instance, considering a computer architecture where preparation and processing zones are spatially separated \cite{kie02}, logically irreversible transformations, such as the reset-to-one operation, could be  performed locally at no energetic cost and without dissipating any heat. The thermodynamic cost associated with the generation of the initial nonequilibrium  state would be restricted to the remote preparation zone before the bit is transferred to the processing area. Remarkably, while information erasure can never be performed  for free, going away from equilibrium  enables  local dissipation-free  logically irreversible reset. 

\section{Supplemental Material}

\subsection{Experimental implementation}

\begin{figure}[h!]
\includegraphics[width=0.49\textwidth]{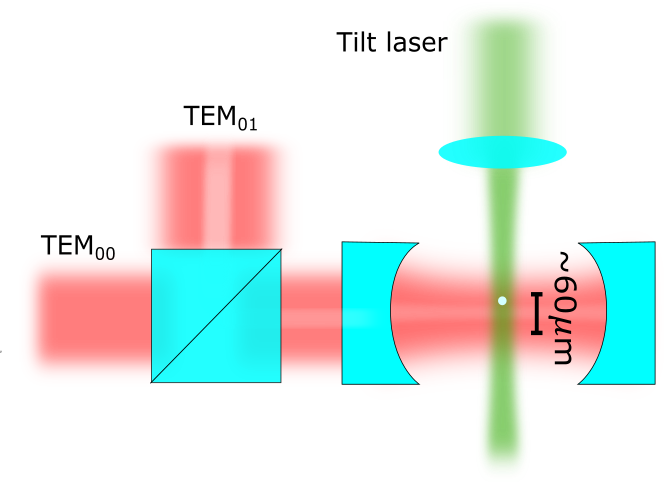}
\caption{Schematic of the proposed experimental setup. A TEM$_{00}$ and a TEM$_{01}$ mode are overlapped on a beam splitter and their frequencies are locked to the cavity resonance of the respective intra-cavity mode. For appropriate power ratios this results in a radial double-well potential for an optically trapped nanoparticle. Along the cavity axis, the particle is tightly confined in the standing-wave lattice. Orthogonally to the cavity axis an off-resonant tilt laser allows to exert radiation pressure on the trapped particle, effectively tilting the double-well potential. Controlling the power of the TEM$_{01}$ laser and the tilt laser enables the implementation of the experimental protocols as discussed in the main text. For the experimental parameters chosen to demonstrate the feasibility of the approach, the cavity beam waist is 43 micrometer. Thus, the potential minima of the two wells are approximately 60 micrometers apart. In addition, the power scattered by the particle, assuming it is from pure silica, is substantial, exceeding 1 microwatt. The motion of the particle can hence be resolved with a simple imaging system from which we can derive the particle trajectories, necessary to determine work and heat. }
\label{fig:setup}
\end{figure}

\textbf{Setup.}
A schematic of the proposed experimental setup is shown in Fig.~\ref{fig:setup}. A double-well potential is formed by the overlap of a TEM$_{00}$ and a TEM$_{01}$ mode. In this potential a silica nanoparticle with a radius of 127 nm is trapped. To achieve a sufficient trap depth, the two modes are power enhanced by an optical cavity. The cavity is mounted in a vacuum environment of approximately $0.5$~mbar pressure. A second laser field, in orthogonal direction, is aligned such that it can exert an additional force on the nanoparticle, effectively tilting the potential for the reset operation. The relevant experimental parameters for implementing the erasure protocol are summarized in Tab.~\ref{tab1} (see Ref.~\cite{kie13}).

The relationship of these experimental parameters to the parameters of the double-well potential used in the numerical simulations is given by the following equations, see Refs.~\cite{Zem08, Har96}:
\begin{widetext}
\begin{equation}
a = \frac{6 P_{00} V}{\pi C W^2} \text{Re}\left( \frac{\epsilon -1}{\epsilon +2} \right), \qquad
b = \frac{48 P_{01} V}{\pi C W^4}\text{Re}\left( \frac{\epsilon -1}{\epsilon +2} \right), \qquad
c = \frac{4}{W^2}, \qquad
A = \frac{256 \pi^4 r^6}{3 C W^2  \lambda^4} \left| \frac{(\epsilon-1)}{(\epsilon+2)} \right|^2 P_{tilt}.
\end{equation}
\end{widetext}
In this conversion the power used before and after the protocol ($g(0)=h(0)=1, f(0)=0$) for each of the three modes enters, with $P_{00}$ ($P_{01}$) the \textit{intracavity power} for the TEM$_{00}$ (TEM$_{01}$) mode and $P_{tilt}$ for the orthogonally directed tilt laser. $W$ the cavity beam waist, $V$ the volume of the nanosphere, $\epsilon$ its relative permeability, and $C$ the velocity of light. \\
The experimental control parameters that allow the implementation of  the initialization step and the protocols for $h(t)$, $g(t)$ and $f(t)$ (Fig.~2 in the main text) are simply the powers $h(t) \cdot P_{00}$ of the TEM$_{00}$ mode, $g(t) \cdot P_{01}$ of the TEM$_{01}$ mode and $f(t) \cdot P_{tilt}$ of the tilt laser, respectively. Note that this requires to go beyond $P_{00}$ and $P_{01}$ in power for the two intra-cavity modes in the initialization step as well as in the initial step of the non-equilibrium erasure protocol. Here, both modes are increased in power by the same amount in order to avoid a reshaping of the potential during initialization. The higher this power increase is in the initialization, the higher is the non-equilibrium parameter $\epsilon$ of the state before the reset protocol. While we do not expect this to pose a challenge for the parameter regime we chose in our analysis (up to $g=h=9.5$), this will set a limit on how far from equilibrium the initial states can be.
\\
The friction coefficient in the Langevin equation (Eq.~(3) in the main text) depends both on the radius $r$ of the nanoparticle and on the pressure $p$. It is given by \cite{li11,Bri}:
\begin{widetext}
\begin{equation}
\begin{split}
\gamma(p,r) = \frac{9\pi \eta}{2\pi r^2\rho_b}\frac{0.619}{0.619+L(p,r)}
\left(1+\frac{0.31L(p,r)}{0.785 + 1.152L(p,r)+L(p,r)^2}\right), \\
\end{split}
\end{equation}
\end{widetext}
with the Knudsen number $L(p,r) = 300k_B /(\sqrt{2}\pi r p d_{air}^2)$, the effective diameter of the air molecules $d_{air}=0.37$~nm, the density of the silica nanosphere $\rho_b = 2100 \text{ kg/m}^3$ and its radius $r$=127~nm.\\

\textbf{Experimental timescales.}
In order to implement the protocols as described in Eqs.~(5)-(8) the experimental setup needs to allow sufficiently fast changes in the optical beam power. For the intracavity modes this is limited by the cavity linewidth $\kappa=180$~kHz and for the tilt laser by the bandwidth of an acousto optic modulator (AOM) used to modulate the laser power. 

For the particular parameters used in the simulations for figures 3-5 in the main text the fastest required time scales are as follows.\\

\begin{itemize}
\item{Initial step of nonequilibrium protocol\\ This step requires a modulation of the power of the TEM$_{01}$ mode on a timescale much shorter than $\tau_{\epsilon}$. The minimal value used in the simulations is $\tau_{\epsilon}^{min}=0.17$~ms. This is compatible with the cavity linewidth.}
\item{Reset protocol: $g(t)$\\ The actual reset protocol requires a modulation of the power of the TEM$_{01}$ mode on the timescale $\tau-t_0$. The minimal value used in the simulations is $(\tau-t_0)^{min}=26.5$~ms. This is compatible with the cavity linewidth.}
\item{Reset protocol: $f(t)$\\ The reset protocol requires a modulation of the power of the tilt laser on a time scale given by the sharp edge of the sawtooth function $f(t)$ as shown in figure 2b in the main text. This slope occurs over the time $0.3 \tau$, which corresponds to $7.95$~ms for the fasted protocol used. This is compatible with the bandwidth of a typical AOM.}
\end{itemize}

Note that the state preparation in the simulation is performed by an instantaneous switch of the power of the TEM$_{01}$ mode.  Experimentally, this can be implemented in good approximation, as the cavity decay time is shorter than the oscillation period of the particle in one well (see also next section). \\

Altogether, the experimental limitations on the speed of power modulation are expected to allow for the experimental demonstration of the protocols simulated in this work.

\subsection{Numerical simulations}
\begin{table}[h]
\begin{tabular}{|c|c|c|c|c|c|c|}
\hline 
$a$ [J] & $b$ [${\text{J}}/{\text{m}^2}$] & $c$ [${1}/{\text{m}^2}$] & $A$ [N] &$\gamma$ [${1}/{\text{s}}]$&$m$ [kg] &$T$ [K]\\ 
\hline 
$1.66\cdot 10^{-18}$&$8.74\cdot 10^{-9}$&$2.16\cdot 10^9$&$6.7\cdot 10^{-15}$&1361&$1.80\cdot 10^{-17}$&293\\
\hline 
\end{tabular} 
\caption{Parameters used in the numerical simulations. These are derived from the following experimental parameters: $P_{00}=210~\text{W},~P_{01}=255~\text{W},~P_{tilt} = 411~\text{mW},~\epsilon = 2.1,~r = 127 \text{nm}$. Further, the cavity beam waist is $W = 43~\mu\text{m}$, the tilt laser beam waist $W_{tilt}=20~\mu\text{m}$. Both lasers have a wavelength of $1064$~nm. The environmental pressure is $p \approx 0.5$~mbar.} 
\label{tab1}
\end{table}
We solve the underdamped Langevin equation with the experimental parameters of Tab.~I using a 4th-order Runge-Kutta algorithm, evaluating the white noise once per step. We compute the work along single trajectories as $\tilde W = \int  dt\, {\partial V(x,t)}/{\partial t}$ \cite{dil09} and the heat from the first law $\tilde Q= \Delta \tilde U - \tilde W$, with the internal energy $\tilde U = p^2/(2m) + V(x)$. Ensemble averages are calculated over 200 000 trajectories. The initial nonequilibrium distribution is taken to be,
\begin{equation}
\rho(x,p,0) = \frac{1}{Z_{b-\varepsilon}} \exp\left[-\beta \left(\frac{p^2}{2m}+ V_{b-\varepsilon}(x)\right)\right],
\end{equation}
where $V_{b-\varepsilon}(x)$ is the double-well potential  at $t=0$ with $b$ replaced by $b-\varepsilon$  or $a,b$ replaced like in (8) in the main text; $Z_{b-\varepsilon}$ is the normalization constant and $\beta=1/(kT)$ is the inverse temperature.
 The starting point of the simulations is randomly  generated according to this distribution. The step-size is chosen by considering a characteristic time of the double-well potential given by  the inverse of its harmonically approximated frequency,
\begin{equation}
\omega = \sqrt{\frac{2}{m}(b-a c)e^{-(1-\frac{ac}{b})}}.
\end{equation}
 We find the value $\tau_0 = 1/\omega \simeq 50~\mu$s for the considered experimental  parameters. We implement a step size of $\Delta t = 0.2 \tau_0$ for the most part of the simulations and a smaller step size of $\Delta t = 0.001\tau_0$ during  the initial rapid change of the potential for  $0< t <\tau_\varepsilon$. We additionally use the respective values $\tau_\varepsilon \simeq 0.17$ ms and $B_\varepsilon = (0.097, 0.135) $ depending on the initial nonequilibrium state, as well as $t_1 = 0.25 \tau$, $t_2 = 0.7\tau$,  $A=6.71\cdot 10^{-15}$N and $B= 0.8936$ for the work improvement nonequilibrium. For the heat improvement nonequilibrium case, the factors are mostly the same, just with  $\tau_\varepsilon =22.5~ms,~ b'=9.5 b,~a'= 9.52 a$. The first sudden jump of $a,~b$ to $a',~b'$ is done in approximately the same time as in the work modification case, or the initialization step, $\tau_1 \simeq 0.17 $ ms.


\begin{thebibliography}{99}
\bibitem{com1} EPROM is an erasable programmable read-only memory  and an EEPROM is an electrically erasable programmable read-only memory \cite{sta15}.
\end{thebibliography}
\end{document}